\documentclass[a4paper,10pt,5p]{elsarticle}

\usepackage{natbib}
\usepackage{graphicx}
\usepackage{subfig}
\usepackage{physics}
\usepackage{feynmf}
\usepackage{amsmath}
\usepackage{comment} 
\usepackage[bottom]{footmisc}

\usepackage[english]{babel}
\usepackage{graphicx}
\usepackage{amsmath, amssymb}
\usepackage{dcolumn}
\usepackage{braket}
\usepackage{amssymb}
\usepackage{amsmath}
\usepackage{lineno,hyperref}  

\journal{Journal of \LaTeX\ Templates}

\newcommand{\be}{\begin{equation}} 
\newcommand{\ee}{\end{equation}}
\newcommand{\bq}{\begin{eqnarray}}
\newcommand{\eq}{\end{eqnarray}}
\newcommand{\ba}{\begin{align}}
\newcommand{\ea}{\end{align}}

\begin{document}

\begin{frontmatter}

\title{Momentum correlations of scattered particles in quantum field theory: one-loop entanglement generation}

\author[BH]{Ricardo Faleiro\corref{cor1}}
\ead{ricardoandremiguel@hotmail.com }
\author[BH]{Rafael Pav\~ao}
\author[MS]{Helder Alexander}
\author[BH]{Brigitte Hiller}
\author[BH]{Alex Blin} 
\author[MS]{Marcos Sampaio}
 
 \cortext[cor1]{Corresponding author}
 
 \address[BH]{CFisUC -- Department of Physics\\ University of Coimbra, 3004
-516 Coimbra, Portugal}

\address[MS]{Universidade Federal de Minas Gerais -- Departamento de F\'{\i}sica -- ICEx \\ P.O. BOX 702,
30.161-970, Belo Horizonte MG -- Brazil}

\begin{abstract}
We compute the entanglement entropy variation between initial (separable or entangled in the momenta) and final states $\Delta S_E$ in an elastic scattering of a bipartite system composed by two interacting scalar particles. We perform a quantum field theoretical calculation to one loop order and verify that $\Delta S_E $ changes as we vary the energy of incoming particles and the coupling strength in a non-trivial way.
\end{abstract}


\end{frontmatter}


\section{Introduction}

Relativistic aspects of quantum entanglement have challenged physicists for more than eighty years now, the milestone being the famous Einstein-Podoslky-Rosen (EPR) gedankenexperiment \cite{EPR}. They concluded that either  single particle entanglement was impossible or the quantum mechanical description of reality was incomplete, which in turn was refuted by Bohr. Thirty years after the EPR paper, J. Bell \cite{Bell} established an inequality  whose violation excludes local realistic theories and validates a spooky action at a distance. Such mathematical formulation has paved the way for Bell test experiments which settle the quantum theory debate between Einstein and Bohr. Seminal experiments carried out at Orsay in 1982 by Aspect, Grangier, Roger and  Dalibard \cite{Aspect} showed a  violation of Bell's inequalities using calcium atoms excited to a particular state, from which the atoms decay by emitting two photons in opposite directions entangled in polarization. Such a pair of entangled photons should be considered as a global, inseparable quantum system. The Aspect experiments show fairly conclusively that quantum physics is non-local.  Recently, Hensen and collaborators \cite{Hensen} conducted efficient measurements of entangled spins in diamonds with a spatial separation of 1.3 kilometres claiming absence of detection loopholes. They tested the CHSH Bell's inequality \cite{CHSH},$S\le 2$ and found $S = 2.42 \pm 0.20$.

Whilst quantum information was originally formulated in terms of non-relativistic quantum mechanics, recent years have seen increasing research interest in placing quantum information within the more fundamental framework of quantum field theory. Relativistic quantum information aims to understand the relationship between special and general relativity and quantum information theory. Quantum entanglement bits (e-bits) are key resources in quantum communication and quantum computation.  Relativistic quantum information plays a key role in studying quantum cryptography, quantum teleportation, quantum computation and quantum metrology \cite{RQIA} both in inertial and noninertial frames. For instance, in  \cite{Friis}  it was pointed out  that gravity or noinertial motion may serve to enhance quantum information protocols. Questions such as how different partitions
of momentum/spin entanglement of relativistic particles or Bell inequalities behave under Lorentz transformations have become important \cite{Bertlmann}. Quantum entanglement also serves as a tool to cosmology. In the early universe, the energy content was largely dominated by highly entangled quantum field background \cite{Menicucci}. Even though experimental evidence show that primordial perturbations have undergone quantum-to-classical transition by some decoherence mechanism, some quantum correlations could in principle linger, in the case of weakly interacting fields, and encode information about the evolution of the universe \cite{Nambu}-\cite{Gustavo}. The theoretical framework to study such phenomena is quantum field theory in curved backgrounds \cite{BF}. The propagation of quantum fields in expanding spacetimes leads to spontaneous creation of pairs of particles with opposite momenta  building up nonlocal quantum correlations. In \cite{Fuentes0} and \cite{Helder2}, the entanglement of quantum scalar field modes of opposite momenta was shown to contain  information about the cosmic parameters characterizing the spacetime expansion. A quantum teleportation protocol for field modes in an expanding spacetime was studied in \cite{Helder1}.

In particular, the study of relativistic scattering when one has access to a particular subset of states in the context of quantum information theory has received a lot of attention. It is rigorously formulated in the framework of Dyson's S-matrix  in relativistic quantum field theory \cite{Fujikawa}. Therefore, entanglement generation in particle decays as well as the variation of entanglement from an incoming to an outgoing state under a quantum field theoretical interaction can be derived in a complete quantum relativistic framework. There exist a plethora of applications of scattering and entanglement generation.  For example in \cite{Yongram} it is studied the violation of Bell's inequalities in polarization correlations in the Standard Model \cite{Yongram} and  in \cite{Seki1} they analyse the ``EPR=ER" conjecture of Maldacena and Susskind in a four gluon scattering represented by open strings. The generation and degree of entanglement in fermionic scattering within quantum electrodynamics was analysed in \cite{Pachos} and a study of entropy variation between  initial and final states asymptotic states to leading order in perturbation theory appears in \cite{Seki2}. An interesting application
appears in \cite{Ratzel}: photon-photon scattering via the QED box diagram. Such a process has a  small cross section and is expected to be measured at the Large Hadron Collider \cite{LHC}. In \cite{Ratzel} they concluded that for a low energy regime that the differential cross section can be written as a function of the degree of entanglement of the incoming photons. An enhancement of the cross section was observed for photons prepared in a symmetric Bell state in their polarizations as compared with the factorized state.

In this work we intend to provide a detailed analysis of entanglement generation  and entropy variation in the scattering of interacting scalar particles in a fully quantum field theoretical framework. Using S-matrix techniques, we derive the final state to one loop order in perturbation theory and express the entropy variation of the reduced state as a function of the degree of entanglement of the initial states. The correlations between the parties show explicit dependence on the speed (energy) between the colliding particles.

This paper is organized as follows:
After presenting the model an the notation in Section 2, we discuss the momentum correlations in the scattering process, in the context of van Neumann's criterion of entanglement in Section 3. Then we obtain in the center of mass frame the tree level and one-loop expressions for the entanglement entropy in section 4, and obtain graphical representations of ${\Delta S}_E$; in sections 4.1 and 4.2 for  the case of an initial separable state, and  in section 4.3 for an initial state which is entangled in momenta. Conclusions and perspectives are presented in section 5.

\section{The model}

Consider an elastic scattering of two kinds of scalar particles $A$ and $B$ in 2-particle Fock space. In quantum field theory such a process can be described by a complex self-interacting scalar field. As we will study this process to one loop order, we may assume unitarity consistently within perturbation theory. The asymptotic (factorized) in-state is
 \begin{equation}
\ket{\Psi}_{in} = \ket{\vec{p_1},\vec{p_2}}= \ket{\vec{p_1}}_A \otimes \ket{\vec{p_2}}_B,
\end{equation}
with,
\begin{equation}
\ket{\vec{p_1}}_A = \sqrt{E_{\vec{p_1}}}c^{\dagger}_{\vec{p_1}}\ket{0}_A  \textup{and} \
\ket{\vec{p_2}}_B = \sqrt{E_{\vec{p_2}}}c^{\dagger}_{\vec{p_2}}\ket{0}_B
\end{equation}

$ c^{\dagger}_{\vec{p}} (c_{\vec{p}}) $ is the creation operator of a mode $\vec{p}$ for  A(lice) defined via the field operator expansion, 
\begin{equation}
\hat{{\phi}}_A(x)= \int \frac{d^3\vec{p_1}}{(2 \pi)^3} \frac{1}{{2E_{\vec{p_1}}}} \left({c}_{\vec{p_1}} e^{-i\vec{p_1}\cdot \vec{x}} +{c^{\dagger}_{\vec{p_1}} e^{i\vec{p_1}\cdot \vec{x}}} \right),
\end{equation}
and similarly for B(ob) with $A \rightarrow B$ and $\vec{p_1} \rightarrow \vec{p_2}$. The commutation relation is normalized such that
\begin{equation*}
[c_{\vec{k}},c^{\dagger}_{\vec{l}}]=2E_{\vec{k}}(2\pi)^3 \delta^{(3)} (\vec{k}-\vec{l}).
\end{equation*}
\be
\langle \vec{q}| \vec{p} \rangle = 2 E_{\vec{q}} (2 \pi)^3 \delta^{(3)} (\vec{q} - \vec{p})
\label{eqN}
\ee

Consider an elastic  scattering $\phi_A \phi_B \rightarrow  \phi_A  \phi_B$, governed by the Scattering Matrix ( $\hat{S}$ ) which is defined as
\begin{equation}
\hat{S}= \mathbf{1} + i\hat{T}.
\label{smatrix}
\end{equation}

We consider the $ \lambda \phi^4$ model for a complex field, $\phi = \phi_A + i \phi_B $ defined by the action,
\begin{equation}
 A = - \int{d^4x}(\partial_{\mu}\phi^*\partial^{\mu}\phi-m^2\phi^*\phi-\frac{\lambda}{4!}(\phi^*\phi)^2.
\end{equation}
In terms of real fields $A$ and $B$ such that $\phi = 1/\sqrt{2} (\phi_A + i \phi_B)$ the action reads
\bq
A &=& - \int{d^4x}\frac{1}{2}\partial_{\mu}\phi_A\partial^{\mu}\phi_A+\frac{1}{2}\partial_{\mu}\phi_B\partial^{\mu}\phi_B \nonumber \\ &-& \frac{m^2}{2}(\phi^2_A+\phi^2_B) - \frac{\lambda}{4!}(\phi^2_A+\phi^2_B)^2.
\eq 
In Tree level we have,

\begin{figure}[ht!]
\centering
\includegraphics[scale=0.25]{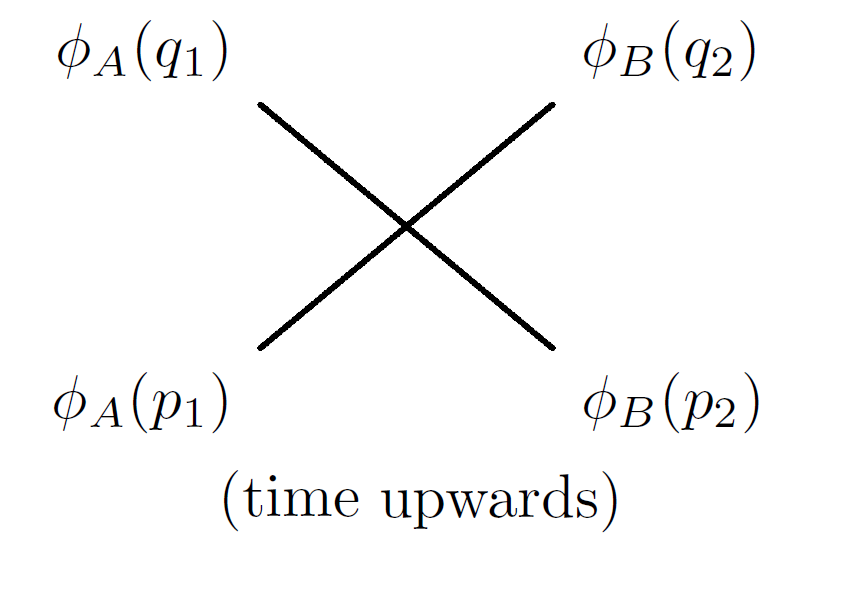}
\caption{Tree-level interaction}
\label{figtree}
\end{figure}

\begin{figure}[ht!]
\includegraphics[scale=0.32]{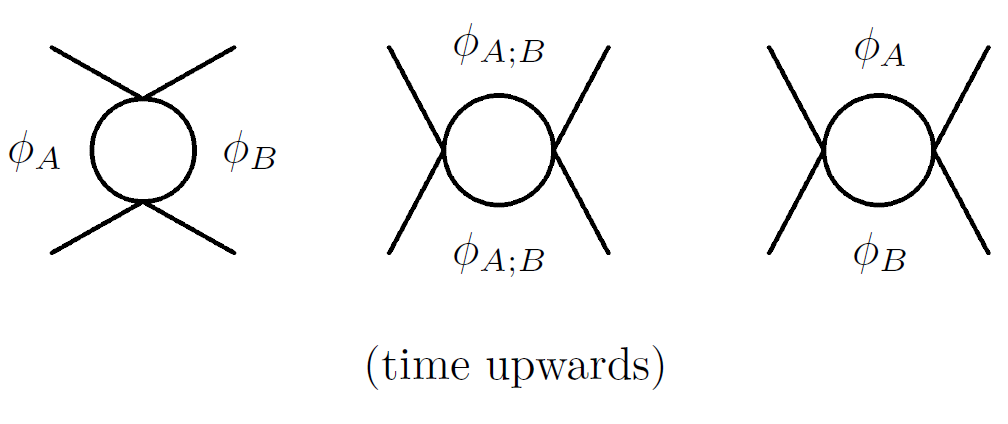}
\caption{These the only three topologically different 1-loop diagrams that respect the coupling term in the Lagrangian. They are the s-channel, t-channel and u-channel, respectively.}
\label{fig1loop}
\end{figure}

With 1-loop corrections we add the contributions shown in fig.\ref{fig1loop} , with the same asymptotic states as in fig.\ref{figtree},

\pagebreak
We shall calculate the reduced density matrix for Alice in order to calculate the von Neumann entropy difference between the initial and final states. First we do this in Tree level, fig.\ref{figtree} and then add 1-loop corrections, represented by the  s, t and u-channels seen in fig.\ref{fig1loop}.

As usual we define the scattering amplitude $\mathcal{M}$ as the matrix element
\bq
\left <  \right. \vec{q_1}\; \vec{q_2} \left. \right|i\hat{T} \left. \right| \vec{p_1}\;\vec{p_2} \left. \right> &=& (2\pi)^4\delta^{(4)} (p_1+p_2-q_1-q_2) \nonumber \\ &\times & i \mathcal{M}_{(p_1,p_2 \rightarrow q_1,q_2)},
\label{mmatrix}
\eq 
which is evaluated as
 \begin{equation}
 i \mathcal{M} =-2i\lambda_R - i(\frac{\lambda_R}{4\pi})^2 \mathcal{F}(u,t,s)+\mathcal{O}(\lambda^3),
 \label{MF}
\end{equation}
 with $\mathcal{F}(u,t,s)= 4 (3 G(t) + 2 G(u) + G(s) + 2)$, $s=(p_1 + p_2)^2$, $t=(p_1 - q_1 )^2$ and $u=(p_1 - q_2)^2$ being the  Mandelstam variables and the  function $G(x)$ stands for
\begin{equation}
G(x)=-2+\sqrt{1-\frac{4m^2}{x}} \ln {(\frac{\sqrt{1-\frac{4m^2}{x}}+1}{\sqrt{1-\frac{4m^2}{x}}-1})},
\label{EG}
\end{equation}
with $x=\{t,u,s\}$. We have adopted a renormalization condition where $\mathcal{M} = - 2i \lambda_R$ at $s =4 m^2$ and $t = u = 0$ in order to define the renormalized coupling constant  $\lambda_R$, hereafter simply denoted by $\lambda$.

\section{Alice's reduced density matrix}
\subsection{\large Density operators and the Von Neumann Entropy of Entanglement}


Von Neumann's entropy of entanglement is known to be an unequivocal measure for bi-partite entanglement. With $\rho_{AB}$ denoting the density operator present in the composite space of Alice and Bob, the \textit{Entropy of Entanglement} between Alice's and Bob's systems is, by definition
\begin{equation}\label{SE}
S_{E} = -Tr\left( \rho_A \log\rho_A\right),
\end{equation}
where $\rho_A$ is Alice's reduced density operator
\begin{equation}
\label{trace}
    \rho_A= Tr_{B} (\rho_{AB})= \int^{+\infty}_{-\infty}  \frac{d^3\vec{n}}{(2 \pi)^3} \frac{\left . \right < \vec{n}|_{B}\;\rho_{AB}\;\left. | \vec{n}\right >_{B}}{2E_{\vec{n}}}.
\end{equation}
In the case of a diagonal operator 
\begin{equation}
  S_E = -\sum_{n=1}^{\infty} ( \rho_{n}) \log(\rho_{n})
 \end{equation}
$\rho_n$ represents the $n$th element of the diagonal, with the density matrix elements in a momentum basis given by
\begin{equation}
\label{element}
\rho_{nl} =
\frac{\left. \right <\vec{n}\left. \right|{{\rho}_{A}}\left. \right|\vec{l} \left. \right>}{\sqrt{2E_{\vec{n}} L^3}\sqrt{2E_{\vec{l}} L^3}},
\end{equation} where $L^3$ is the phase space volume defined as $L^3=(2\pi \delta(0))^3$ which, just as the squared energy delta function, will be eliminated with a proper normalization, further on. We are interested in studying its variation during a scattering process, 
\begin{equation}
    \label{var}\Delta S_{E} = S_{E\;(out)}-S_{E\;(in)}.
    \end{equation}
The process to calculate it will be the following, 

\begin{equation}
\label{recipe}
    \ket{\Psi}_{in} \rightarrow \rho^{(in)}_{AB} \rightarrow \rho^{(in)}_{A} \rightarrow S_{E\;(in)}
\end{equation}

\[\!\!\!\!\!\!\!\!\! \ket{\Psi}_{out} \rightarrow \rho^{(out)}_{AB} \rightarrow \rho^{(out)}_{A} \rightarrow S_{E\;(out)}\]

and obviously in the end use (eq.\ref{var}).

\subsection{\large When the \textit"{in state"} is separable }

In this case we define the state as
\begin{equation}
\ket{\Psi}_{in} = \ket{\vec{p_1},\vec{p_2}}= \ket{\vec{p_1}}_A \otimes \ket{\vec{p_2}}_B
,
\end{equation}


Following the recipe of (eq.\ref{recipe}), one realizes that since the initial state is separable, by definition, the entropy of entanglement vanishes before the collision, and (eq.\ref{var}) becomes just $S_{E\;(out)}$.


We shall work in the center-of-mass (CM) frame in which the in-state is given by $\ket{\vec{p_1},-\vec{p_1}}$ and a general out-state
can be written as $$ \ket{(\vec{q_1},\vec{q_2})} = c_{\vec{p_1}}\ket{(\vec{p_1},-\vec{p_1})} + \sum_{\forall \vec k \neq {\vec{p_1}}} c_{\vec{k}}|(\vec{k},-\vec{k}) \rangle   ,$$ the parenthesis meaning that the out-state is generally entangled in the momenta. It is calculated using (\ref{smatrix}) and (\ref{mmatrix}) as
 \begin{equation}
 \label{psii}
     \ket{\Psi}_{out}=
     |\vec{p_1}\;\vec{p_2}\left. \right> +\!\!\!\!\!\!\!\!\iint\limits_{\vec{q_1}\neq \vec{p_1};\\ \vec{q_2}\neq \vec{p_2}}  \frac{ |\vec{q_1}\;\vec{q_2} \left. \right>\left <  \right. \vec{q_1}\; \vec{q_2} \left. \right|i \hat{T} \left. \right| \vec{p_1}\;\vec{p_2} \left. \right>}{2E_{\vec{q_1}}{2E_{\vec{q_2}}}}, \\ 
    \end{equation}
\bq
 &=& |\vec{p_1}\;\vec{p_2}\left. \right> + \iint\limits_{\vec{q_1}\neq \vec{p_1}\vec{q_2}\neq \vec{p_2}}  \frac{ (2\pi)^4\delta^{(4)} (p_1+p_2-q_1-q_2)}{2E_{\vec{q_1}}{2E_{\vec{q_2}}}} \nonumber  \\ 
&\times& i\mathcal{M}_{(p_1 p_2 \rightarrow q_1 q_2)}|\vec{q_1}\;\vec{q_2} \left. \right>
\eq
with $\int_{\vec{p}} \equiv \int d^3\vec{p}/(2\pi)^3$ and we normalize the inner product as 
$ \langle \vec{q}|\vec{p} \rangle =2E_{\vec{q}}(2\pi)^3 \delta^{(3)} (\vec{q}-\vec{p})$. Moreover we separate the $\delta^{(4)}$ into $\delta^{(3)} (\vec{p_1}+\vec{p_2}-\vec{q_1}-\vec{q_2})\delta(E_{\vec{p_1}}+E_{\vec{p_2}}-E_{\vec{q_1}}-E_{\vec{q_2}})$. Consequently,
\begin{equation}
     \ket{\Psi}_{out}=
     |\vec{p_1}\;\vec{p_2} \left. \right>+ \!\!\!\int\limits_{\vec{q_1}\neq \vec{p_1}} \frac{ |\vec{q_1},\;\vec{p_1}+\vec{p_2}-\vec{q_1} \left. \right>[(2\pi)\delta (E)] i \mathcal{M}}{2E_{\vec{q_1}}2E_{\vec{p_1}+\vec{p_2}-\vec{q_1}} },
     \label{psiout}
    \end{equation}
in which  $\delta (E)$  stands for $\delta (E_{\vec{p_1}}+E_{\vec{p_2}}-E_{\vec{q_1}}-E_{\vec{p_1}+\vec{p_2}-\vec{q_1}})$ and 
$\mathcal{M}$ is evaluated with four-momentum \\ $ q_2 = (E_{\vec{p_1}+\vec{p_2}-\vec{q_1}},\vec{p_1}+\vec{p_2}-\vec{q_1})$. From (\ref{psiout}) we can evaluate $$\hat{\rho}_{out} = |\Psi \rangle_{out} \,\,  {}_{out} \langle \Psi |, $$
as well as the reduced density matrix for Alice, integrating out the momenta of Bob, $\vec{p_2}$. Explicitly,

\begin{equation}
\hat{\rho}_{A}= \left. \right|\vec{p_1} \left. \right>\left. \right <\vec{p_1}\left. \right|(2E_{\vec{p_2}}L^3)+\int\limits_{\vec{q}\neq \vec{p_1}} \frac{\left. \right |\vec{q} \left. \right>\left. \right <\vec{q} \left .\right |[(2\pi)\delta (E)]^2 |\mathcal{M}|^2}{(2E_{\vec{q}})^2 2E_{\vec{p_1}+\vec{p_2}-\vec{q}}  }.
\end{equation}
 In order to compute the reduced density matrix let us split $ \mathcal{F}(t,u,s)$ in equation (\ref{MF}) into real and imaginary parts. Hence 

\begin{equation}
\label{M}
|\mathcal{M}|^2 = 4\lambda^2 + 4\lambda(\frac{\lambda}{4\pi})^2\textrm{Re}[\mathcal{F}]+(\frac{\lambda}{4\pi})^4 \big(\textrm{Re}^2[\mathcal{F}]+\textrm{Im}^2[\mathcal{F}]\big)
\end{equation}
which enables us to write
\begin{equation}
    \hat{\rho}_{A}= \left. \right|\vec{p_1} \left. \right>\left. \right <\vec{p_1}\left. \right|(2E_{\vec{p_2}}L^3)+\mathcal{I}_2+\mathcal{I}_3+\mathcal{I}_4 ,
\end{equation}
with 
\begin{equation*}
    \mathcal{I}_2 \equiv \int\limits_{\vec{q}\neq \vec{p_1}}  \frac{\left. \right |\vec{q} \left. \right>\left. \right <\vec{q} \left .\right |[(2\pi)\delta (E)]^2 }{(2E_{\vec{q}})^22E_{\vec{p_1}+\vec{p_2}-\vec{q}}  }(4\lambda^2),
\end{equation*}

\begin{equation*}
    \mathcal{I}_3 \equiv \int\limits_{\vec{q}\neq \vec{p_1}}  \frac{\left. \right |\vec{q} \left. \right>\left. \right <\vec{q} \left .\right |[(2\pi)\delta (E)]^2 }{(2E_{\vec{q}})^22E_{\vec{p_1}+\vec{p_2}-\vec{q}}  }(4\lambda(\frac{\lambda}{4\pi})^2\textrm{Re}[\mathcal{F}]),
\end{equation*}

\begin{equation*}
    \mathcal{I}_4 \equiv \int\limits_{\vec{q}\neq \vec{p_1}}  \frac{\left. \right |\vec{q} \left. \right>\left. \right <\vec{q} \left .\right |[(2\pi)\delta (E)]^2 }{(2E_{\vec{q}})^2 2E_{\vec{p_1}+\vec{p_2}-\vec{q}}  }(\frac{\lambda}{4\pi})^4(\textrm{Re}^2[\mathcal{F}]+\textrm{Im}^2[\mathcal{F}]).
\end{equation*}
We normalize the reduced density operator by demanding that $Tr_{A} (\mathcal{N}\hat{\rho}_{A})= Tr_{A} (\hat{\rho}_{A}^n)= 1$ which amounts to requiring that
\begin{equation}
    \mathcal{N} \int^{+\infty}_{-\infty}  \frac{d^3\vec{n}}{(2 \pi)^3} \frac{\left . \right < \vec{n}|\;\hat{\rho}_{A}\left. | \vec{n}\right >}{2E_{\vec{n}}}=1 .
\end{equation}
Setting 
\begin{equation}
   \int^{+\infty}_{-\infty}  \frac{d^3\vec{n}}{(2 \pi)^3} \frac{\left . \right < \vec{n}|\;\mathcal{I}_i \left. | \vec{n}\right >}{2E_{\vec{n}}}\equiv \langle\mathcal{I}_i\rangle \,\,\, , \,\, i = 2, 3, 4, 
\end{equation} 
yields $$ \mathcal{N} = [(2E_{\vec{p_1}}L^3)(2E_{\vec{p_2}}L^3) + \langle\mathcal{I}_2\rangle + \langle\mathcal{I}_3\rangle + \langle\mathcal{I}_4\rangle ]^{-1}$$ and the {\it{renormalized}} density operator to one loop order formally reads
\be
\label{rn}
\hat{\rho}_A = \frac{|\vec{p_1} \rangle \langle \vec{p_1}| (2 E_{\vec{p_2}} L^3) + \mathcal{I}_2 + \mathcal{I}_3 + \mathcal{I}_4}
{(2 E_{\vec{p_1}} L^3)(2 E_{\vec{p_2}} L^3)+\langle\mathcal{I}_2\rangle + \langle\mathcal{I}_3\rangle + \langle\mathcal{I}_4\rangle},
\ee
where we have dropped the superscript $n$ for brevity. Notice that $\mathcal{I}_3$ and $\mathcal{I}_4$ represent the one loop correction to the scattering amplitude.

\section{Transition amplitude in CM variables}
In the CM frame we have
$E_{\vec{p_1}} = E_{\vec{p_2}} \equiv E_{CM}$,  $\vec{p_1} = - \vec{p_2} \rightarrow |\vec{p_1}| = |\vec{p_2}| \equiv |\vec{p}_{CM}|$ and the energy delta function $\delta (E)$ becomes $ \delta (2E_{CM}-2E_{\vec{q}})$. Hence, the Mandelstam variables in the one loop correction to the transition amplitude in equation (\ref{MF}) read, in the CM coordinates,
\bq
u_{CM} &=& - 2 p^2_{CM}(1+\cos\theta), \nonumber \\
t_{CM} &=& - 2 p^2_{CM}(1-\cos\theta) \,\,\, \mathrm{and} \nonumber \\
s_{CM} &=& 4(m^2 + p^{2}_{CM}),
\label{MCM}
\eq 
$\theta$ being the scattering angle. Using equations (\ref{MF}), (\ref{EG}) and (\ref{MCM}), the real and the imaginary parts of $\mathcal{F}$ become a function of the velocity   and the scattering angle in the CM frame,  ${\cal F}(\vec{p},\theta)$, namely
\begin{equation}
\label{RIF}
\textrm{Re}[\mathcal{F}] = 12G\left(-2p^2\cdot(1+\cos\theta)\right) + 8G\left(-2p^2\cdot(1-\cos\theta)\right)  \end{equation}\[\!\!+ 4 |v_{\vec{p}}| \log{\left(\frac{1+|v_{\vec{p}}|}{1-|v_{\vec{p}}|}\right)};\;\;\;\;\textrm{Im}[\mathcal{F}] = 4\pi (|v_{\vec{p}}|);\]
where we use the shorthand notation $\vec{p}_{CM}\equiv \vec{p}$, and $v_{\vec{p}}$ stands for the CM velocity.

 
 

\subsection{\large Tree-level}
First we calculate  tree-level result for the entanglement entropy, from $|\mathcal{M}|^2= 4\lambda^2$.

For this case eq. (\ref{rn}) reduces to 
\begin{equation}
\label{rhoA}
{\rho_{A}}^{(n)} = \frac{\left. \right|\vec{p} \left. \right>\left.\right <\vec{p}\left. \right|(2E_{\vec{p}}L^3)+\mathcal{I}_2}{(2E_{\vec{p}}L^3)^2 +\langle\mathcal{I}_2\rangle }
\end{equation}\[ = \frac{\left. \right|\vec{p} \left. \right>\left. \right <\vec{p}\left. \right|}{(2E_{\vec{p}}L^3)(1 +\mathcal{A}_{tree})}+  \frac{\mathcal{I}_2}{(2E_{\vec{p}}L^3)^2(1 +\mathcal{A}_{tree})}.\]
where 
\[\mathcal{A}_{tree}= \frac{\langle\mathcal{I}_2\rangle}{(2E_{\vec{p}} L^3)^2}\]  which yields in the CM
\[\mathcal{A}_{tree}=\frac{\lambda^2}{8\pi}\frac{|v_{\vec{p}}|}{(E_{\vec{p}}L)^2}\]

We can now calculate the entropy associated with the state after the collision, by applying (eq.\ref{trace}), or since we know
the matrix is diagonal (eq.\ref{SE}),
\begin{equation}
\label{entropy}
(S_E)_{tree} = -\rho_{p}\log{\rho_{p}}
   - \frac{L^3}{(2\pi)^3}\int^{+\infty}_{-\infty} (\rho_{k}\log\rho_{k})d\vec{k}, k\neq p.
\end{equation}

 We can easily compute $\rho_{p}$ and $\rho_{k}$ by using (eq.\ref{element}) and (eq.\ref{rhoA}), \[\rho_{p} =  \frac{1}{1+\mathcal{A}_{tree}}; \;\;\rho_{k} = \left( \frac{2\pi\delta(2E_{\vec{p}}-2E_{\vec{k}})}{2E_{\vec{k}} 2E_{\vec{p}}L^3}\right)^2 \frac{4\lambda^2}{1+\mathcal{A}_{tree}}.\]

Substituting into (eq.\ref{entropy}) and solving the integral in the second term we have, \begin{equation}
\label{SEtree}
(S_E)_{tree} =
\frac{\log{(1+\mathcal{A}_{tree})}}{1+\mathcal{A}_{tree}} +\lambda^2|v_{\vec{p}}|\frac{\log{\left((4\bar{E}_{\vec{p}}^4(1+\mathcal{A}_{tree}))/\lambda^2 \right)}}{4\bar{E}_{\vec{p}}^2(1+\mathcal{A}_{tree})},
\end{equation}$\bar{E}_{\vec{p}}$ is defined to be ${E}_{\vec{p}}L = \gamma_{\vec{p}}\;\bar{m}$, if $\bar{m}=mL$.

It is worth mentioning that this result is equivalent to the similiar result obtained in \cite{Seki1}.
\vspace{0.5cm}

\subsection{\large 1-Loop corrections}
\vspace{0.5cm}In a similar way one obtains up to 1-Loop, the following normalized reduced density operator
\begin{equation}
\label{normreduced}
{\rho_{A}}^{(n)} = \frac{\left. \right|\vec{p} \left. \right>\left. \right <\vec{p}\left. \right|}{(2E_{\vec{p}}L^3)(1 +\mathcal{A}_{1loop})}+ 
\end{equation}
\begin{equation*}\frac{1}{(1 +\mathcal{A}_{1loop})}\left (
    \frac{\mathcal{I}_2}{{(2E_{\vec{p}}L^3)^2}}+ \frac{\mathcal{I}_3}{{(2E_{\vec{p}}L^3)^2}}+ \frac{\mathcal{I}_4}{{(2E_{\vec{p}}L^3)^2}}\right )
\end{equation*}
 where \[\mathcal{A}_{1loop}=
 \frac{\langle\mathcal{I}_2\rangle}{(2E_{\vec{p}}L^3)^2}+ \frac{\langle\mathcal{I}_3\rangle}{(2E_{\vec{p}}L^3)^2} + \frac{\langle\mathcal{I}_4\rangle}{(2E_{\vec{p}}L^3)^2} .\]
Explicitly the integrals read
\begin{equation}
 \langle\mathcal{I}_{3}\rangle = \frac{L^4}{4\pi} \lambda(\frac{\lambda}{4\pi})^2|v_{\vec{p}}|(-120 + \log{\frac{1/|v_{\vec{p}}| +1}{1/|v_{\vec{p}}|-1}}\left (\frac{40}{|v_{\vec{p}}|}\right  )+\end{equation}\[ + (\frac{10}{v^2_{\vec{p}}}-10)\left (\log{\frac{1/|v_{\vec{p}}| +1}{1/|v_{\vec{p}}|-1}}\right )^2 + 8|v_{\vec{p}}|\log\frac{1+|v_{\vec{p}}|}{1-|v_{\vec{p}}|}),\]
and  
  \begin{equation}
   \langle\mathcal{I}_{4}\rangle=\frac{L^4}{16\pi}(\frac{\lambda}{4\pi})^4 |v_{\vec{p}}|\left( \int^{\pi}_{0} (\textrm{Re}[\mathcal{F}])^2\sin\theta d\theta+ 32\pi^2 v^2_{\vec{p}}\right),
\end{equation}
which is computed numerically.

Again we calculate $\rho_{p}$ and $\rho_{k}$ by using (eq.\ref{element}) and now (eq.\ref{normreduced}) , \[\rho_{p} =  \frac{1}{1+\mathcal{A}_{1loop}}; \;\;\rho_{k} = \left( \frac{2\pi\delta(2E_{\vec{p}}-2E_{\vec{k}})}{2E_{\vec{k}} 2E_{\vec{p}}L^3}\right)^2 \frac{\Omega(\vec{k},\theta^{'})}{1+\mathcal{A}_{1loop}},\]
where $\Omega({\vec{k}},\theta^{'})$
is given by the same expression as (\ref{M}) , with ${\cal F}(\vec{k},\theta^{'})$, see (\ref{RIF}).
Substituting these elements into the entropy (eq.\ref{entropy}) and integrating the deltas, we get  to 1-loop, 
\begin{eqnarray}
\label{SEloop}
&&(S_E)_{1-loop}=    \frac{\log{{(1 +\mathcal{A}_{1loop})}}}{(1 +\mathcal{A}_{1loop})} + \nonumber \\
&&  \frac{|v_{\vec{p}}|/64\pi}{(\bar{E}_{\vec{p}})^2(1+\mathcal{A}_{1loop})} \times \nonumber \\
&&\int^{\pi}_{0} \sin\theta^{'}\Omega(\theta^{'})\log{[\frac{16(\bar{E}_{\vec{p}})^4(1+\mathcal{A}_{1loop})}{\Omega(\theta^{'})}]d\theta^{'}}.
\end{eqnarray}
It should be noticed that if we take (eq.\ref{SEloop}) and take the limiting case when $\mathcal{A}_{1loop}\rightarrow \mathcal{A}_{tree}$ and $\Omega(\theta^{'}) \rightarrow 4\lambda^2$ we reproduce (eq.\ref{SEtree}). 

\subsubsection{Graphical solutions}

We take (eq.\ref{SEloop}), (eq.\ref{SEtree}), and choose $\bar{m}=1$.
All the graphics are of the form $(S_E$ vs $|v_{\vec{p}}|)$. The Tree-level contribution is in orange and the Tree-level with the higher order corrections (up to 1-loop) is in blue.
In fig. \ref{fig:table2} we vary increasingly the value of the coupling $\lambda$, staying in the perturbative regime, i.e. the variations in the maxima of $S_E$ between the  Tree level plus 1-loop corrections and Tree level alone are required to be less than $30\%$.

\begin{figure}[h!]
\centering
    \subfloat{\vspace{1em}
\includegraphics[scale=0.30]{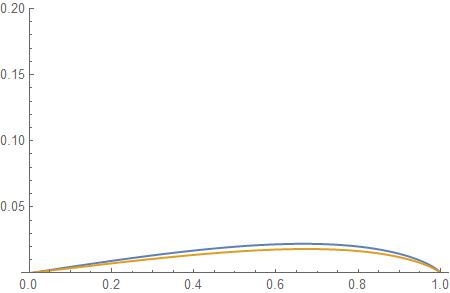}

}
\subfloat{\vspace{1em}
\includegraphics[scale=0.30]{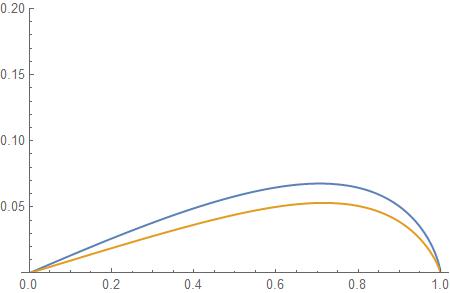}

}
\label{fig:table1}

\end{figure}
\begin{figure}[h!]
\centering
   \subfloat{\vspace{1em}
\includegraphics[scale=0.30]{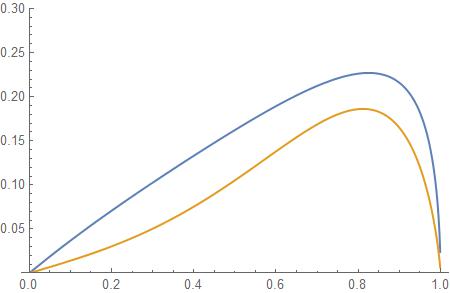}
}
\subfloat{\vspace{1em}
\includegraphics[scale=0.30]{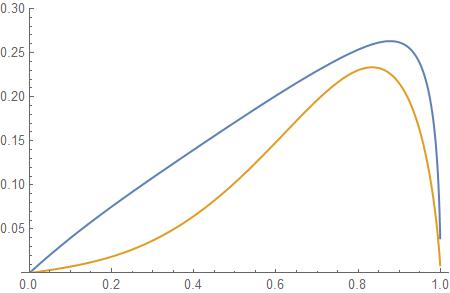}
}
\caption{{$S_E(|v_{\vec{p}}|)$ in Tree-level (orange) and 1-loop (blue), for}$\lambda$=[0.5(top left), 1(top right), 2.5(bottom left), 3(bottom right)]}
\label{fig:table2}
\end{figure}

\begin{figure}[h!]
\centering
\includegraphics[scale=0.5]{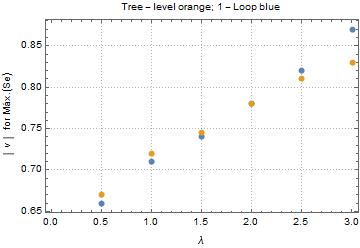}
\caption{$|v_{\vec{p}}|$ vs. $\lambda$}
\label{fig:velocitylinear}

\end{figure}

One sees that the velocity for the maximum value of the entropy increases with $\lambda$, actually it increases in an approximately linear way, as shown in fig. \ref{fig:velocitylinear}. 

One can appreciate that the velocities in the 1-loop corrections  at the maximum  of entropy are only slightly different than the velocities in Tree-level. 

\begin{figure}[h!]
\centering
\includegraphics[scale=0.5]{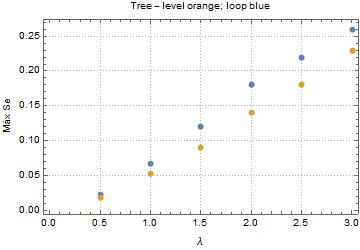}
\caption{$S_E$ vs. $\lambda$}
\label{fig:entropylinear}

\end{figure}

The variation of the entropy maximum as a function of $\lambda$ is shown in fig. \ref{fig:entropylinear}. Note that it is strongly dependent on the value of the coupling, one order of magnitude difference when read at $\lambda=0.5$ and $\lambda=3$.

\subsection{\large When the before state is entangled}

Now, in analogy to a typical Bell state, we propose the following simplest possible  initial entagled state in a momentum basis, already seen from the perspective of the CM,
\begin{eqnarray}
\label{entstate}
&&    \ket{\Phi}_{\textit{in}} = \alpha_p\ket{\vec{p},-\vec{p}} + \alpha_k\left. \right|\!\vec{k}, -\vec{k} \!\left. \right> ,  \; \forall \left \{  \vec{p} \neq \vec{k}\right \}; \nonumber \\
&&   |\alpha_p|^2 +|\alpha_k|^2 = 1 . 
\end{eqnarray}

Following the recipe of (eq. \ref{recipe}), we now calculate the reduced operator and consequently the entropy associated with the initial state. Unlike the previous case it won't be null.

Alice's already normalized operator is given by, 

\begin{equation}
   \rho_{A} = \frac{\ket{\vec{p}}\bra{\vec{p}}}{(L^3)(2E_{\vec{p}})\left( 1 + 1/\mathcal{C}\right)}+\frac{\left. \right|\!\vec{k}\!\left. \right>\left. \right<\!\vec{k}\!\left. \right|}{(L^3)(2E_{\vec{k}})\left( 1 + \mathcal{C}\right)} ,
\end{equation}

where $\mathcal{C}=  \frac{|\alpha_p|^2}{|\alpha_k|^2} \left(\frac{E_{\vec{p}}}{E_{\vec{k}}}\right)^2 =  \frac{|\alpha_p|^2}{|\alpha_k|^2} \left(\frac{1-v_{\vec{k}}^2}{1-v_{\vec{p}}^2}\right)$.

We calculate the elements $\rho_{p}$ and $\rho_{k}$, by using (eq.\ref{element} again and inputting into (eq.\ref{entropy}), we get for the initial state entanglement

\begin{equation}
\label{SE_ENT}
  S_E= \frac{\log{(1+1/\mathcal{C}})}{1+1/\mathcal{C}}+\frac{\log{(1+\mathcal{C}})}{1+\mathcal{C}}
\end{equation}

It can be easily seen that the entropy is maximum for $\mathcal{C}=1$, that represents the case of maximum entanglement. One can appreciate  that for any choice of non null values of $\alpha_p$ and $\alpha_k$ we can get a maximum entangled state for an appropriate choice of  velocities values. This becomes evident if we take the condition for maximum entanglement, \[ \frac{|\alpha_p|^2}{|\alpha_k|^2} \left(\frac{1-v_{\vec{k}}^2}{1-v_{\vec{p}}^2}\right) = 1,\] and rewrite it as,\[( 1-|\alpha_p|^2)(1-v_{\vec{p}}^2) + |\alpha_p|^2(v_{\vec{k}}^2-1) = 0 , \;|\alpha_p|^2\in (0,1)\]

This is the equation of a line segment between two points, one is negative ($v_{\vec{k}}^2-1$) and the other one is positive ($1-v_{\vec{p}}^2$), so regardless what their values are there is always one value of $|\alpha_p|^2$ where the line passes trough the origin, hence the condition of maximum entangled is satisfied. For $|\alpha_p|^2=|\alpha_k|^2=0.5$, this point is  when $|v_{\vec{p}}|=|v_{\vec{k}}|$.

The plot of (eq.\ref{SE_ENT}) as a function of $\mathcal{C}$ is
 
\begin{figure}[h!]
\centering
\includegraphics[scale=0.55]{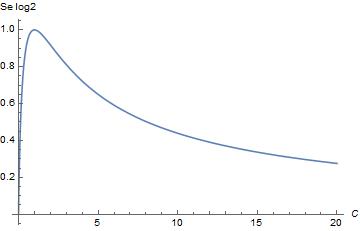}
\caption{$S_E$($\mathcal{C}$). One can notice that for values where $\mathcal{C}$ goes to infinity or to zero the entropy vanishes (separable case), this being equivalent as to making one of the coefficients null.}
\label{fig:S}
\end{figure}
\begin{figure}[h!]
\centering
\includegraphics[scale=0.4]{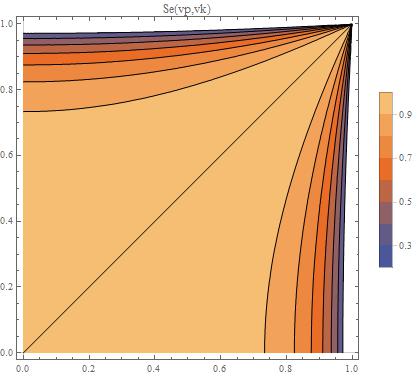}

\caption{ All the points in the diagonal line $|v_{\vec{p}}|= |v_{\vec{k}}|$ correspond to the peak of maximum entropy in fig.\ref{fig:S} when $\mathcal{C}=1$. This is easily understood since for every point of equal velocity when the coefficients are the same the ratio is always 1. }
\label{Spk}
\end{figure}

If instead of the initial two-state entanglement in a continuous momentum variable one would have taken a discrete two parameter state, as a Bell state of spins, the curve would e symmetric around the maximum. 
In fig. \ref{Spk} we display the contour plot of $S_E(|v_{\vec{p}}|,|v_{\vec{k}}|)$, for $|\alpha_p|^2=|\alpha_k|^2=0.5$. 

Now that we have concluded the study of the entropy before the collision we evaluate the entropy after the collision. We compute the sate after the collision by taking  (eq.\ref{entstate}) projecting it into a basis of momenta like in (eq.\ref{psii}) and applying (eq.\ref{smatrix}). Then we can construct the density operator and calculate Alice's reduced density operator, following the steps in the recipe (eq.\ref{recipe}). 

The normalized operator in Tree-level when $|\alpha_p|^2=|\alpha_k|^2=0.5$ takes the form,

\begin{equation}
\rho_{A}= \mathcal{N}(\ket{\vec{p}}\bra{\vec{p}}(2E_{\vec{p}}) +  |\vec{k} \left.\right>\!\!\left.\right<\vec{k}|(2E_{\vec{k}})+
\end{equation}\[+ 4\lambda^2\!\!\!\!\!\!\!\! \int\limits_{\left \{\vec{q}\neq \vec{p}, \vec{k}\right \}}\!\!\!\!\frac{d^3\vec{q}}{(2\pi)^3}\frac{\ket{\vec{q}}\bra{\vec{q}}}{(2E_{\vec{q}})^3 } \left(2\pi {\delta}(2E_{\vec{p}_{CM}} - 2E_{\vec{q}})\right)^2 + \]\[+4\lambda^2 \!\!\!\!\!\!\! \int\limits_{\left \{\vec{q}\neq \vec{p}, \vec{k}\right \}}\!\!\!\!\frac{d^3\vec{q}}{(2\pi)^3}\frac{\ket{\vec{q}}\bra{\vec{q}}}{(2E_{\vec{q}})^3 } \left(2\pi {\delta}(2E_{\vec{k}} - 2E_{\vec{q}})\right)^2) .\]
 Where $\mathcal{N}$ is calculated to be, 
 
 \begin{equation*}
     \mathcal{N}= \frac{1}{(2E_{\vec{p}}L^3)^2 + (2E_{\vec{k}}L^3)^2 + \frac{\lambda^2}{2\pi}L^4(|v_{\vec{p}}|+|v_{\vec{k}}|)}
\end{equation*}

Now the entropy has one more element adding to the remaining infinite amount, such that the entropy becomes, 

\begin{equation}
\label{entSE}
     \left(S_E\right)_{out} =  -\rho_{p}\log{\rho_{p}}
 -\rho_{k}\log{\rho_{k}}- \frac{L^3}{(2\pi)^3}\int_{-\infty}^{+\infty}\rho_{n}\log{\rho_{n}}dn\end{equation}
 
 Finding the elements in the usual way, 
  \[   \rho_p= \frac{1}{1 + \frac{1}{\mathcal{C}}+\frac{\lambda^2}{8\pi\bar{E}_{\vec{p}}}(|v_{\vec{p}}|+|v_{\vec{k}}|)}, \]   \[  \rho_{k} = \frac{1}{1 + \mathcal{C}+\frac{\lambda^2}{8\pi\bar{E}_{\vec{k}}}(|v_{\vec{p}}|+|v_{\vec{k}}|)}. \]
  
  And finally, 

\[\rho_n = \frac{(2\pi)^2\left( \delta^2(2E_{\vec{p}}-2E_{\vec{n}})+ \delta^2(2E_{\vec{k}}-2E_{\vec{n}})\right)}{L^6(2E_{\vec{n}})^2\left((2E_{\vec{k}})^2+(2E_{\vec{p}})^2+\frac{\lambda^2}{2\pi L^2}(|v_{\vec{p}}| +|v_{\vec{k}}|)\right)}\]

Inputting this all into (eq.\ref{entSE}) we get the expression for the entropy after the collision, if we subtract (eq.\ref{SE_ENT}) we get it's variation. After all the integrations we get explicitly, 

\begin{equation}
\label{final}
       \Delta S_E = \frac{\log{\left(1 + \frac{1}{\mathcal{C}}+\frac{\lambda^2}{8\pi\bar{E}^2_{\vec{p}}}(|v_{\vec{p}}|+|v_{\vec{k}}|)\right)}}{1 + \frac{1}{\mathcal{C}}+\frac{\lambda^2}{8\pi\bar{E}^2_{\vec{p}}}(|v_{\vec{p}}|+|v_{\vec{k}}|)}
\end{equation}\[+\frac{\log{\left(1 + \mathcal{C}+\frac{\lambda^2}{8\pi\bar{E}^2_{\vec{k}}}(|v_{\vec{p}}|+|v_{\vec{k}}|)\right)}}{1 + \mathcal{C}+\frac{\lambda^2}{8\pi\bar{E}^2_{\vec{k}}}(|v_{\vec{p}}|+|v_{\vec{k}}|)}\]\[+\lambda^2|v_{\vec{p}}|\frac{\log{\left((4\bar{E}^4_{\vec{p}} + 4 \bar{E}^2_{\vec{p}}\bar{E}^2_{\vec{k}}+\bar{E}^2_{\vec{p}}\frac{\lambda^2}{2\pi}(|v_{\vec{p}}|+|v_{\vec{k}}|))/\lambda^2 \right)}}{2\pi\left(4\bar{E}^2_{\vec{p}}+4\bar{E}^2_{\vec{k}}+\frac{\lambda^2}{2\pi}(|v_{\vec{p}}|+|v_{\vec{k}}|)\right)}+\]\[+\lambda^2|v_{\vec{k}}|\frac{\log{\left((4\bar{E}^4_{\vec{k}} + 4 \bar{E}^2_{\vec{p}}\bar{E}^2_{\vec{k}}+\bar{E}^2_{\vec{k}}\frac{\lambda^2}{2\pi}(|v_{\vec{p}}|+|v_{\vec{k}}|))/\lambda^2 \right)}}{2\pi\left(4\bar{E}^2_{\vec{p}}+4\bar{E}^2_{\vec{k}}+\frac{\lambda^2}{2\pi}(|v_{\vec{p}}|+|v_{\vec{k}}|)\right)}\]\[-\left(\frac{\log{(1+1/\mathcal{C}})}{1+1/\mathcal{C}}+\frac{\log{(1+\mathcal{C}})}{1+\mathcal{C}}\right)\]

\subsubsection{Graphical solutions}

Using the same prototype for $\bar{m}=1$, we can find graphical solutions to this expression, 
\begin{figure}[ht!]

    \subfloat{\vspace{2em}
\includegraphics[scale=0.3]{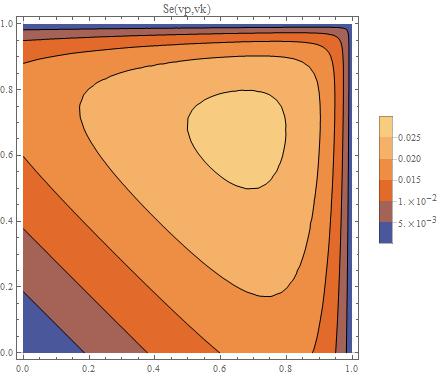}

}
\subfloat{\vspace{2em}
\includegraphics[scale=0.3]{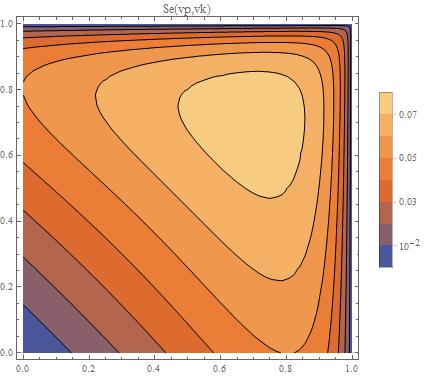}

}
\label{fig:contours}

\end{figure}

\begin{figure}[ht!]
\centering
   \subfloat{\vspace{2em}
\includegraphics[scale=0.3]{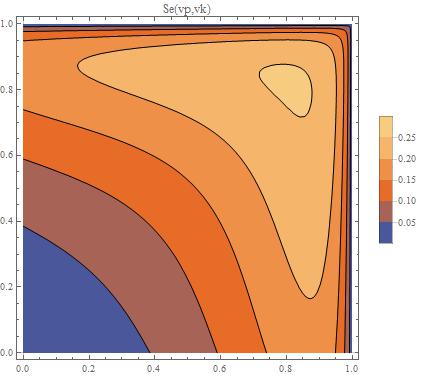}
}
\subfloat{\vspace{2em}
\includegraphics[scale=0.3]{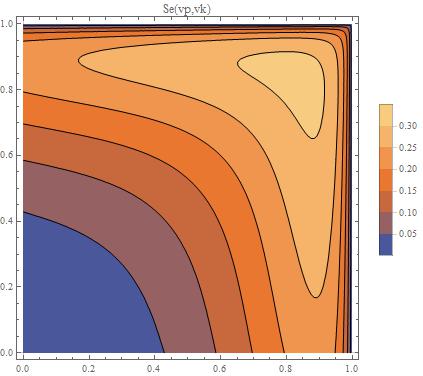}
}
\caption{{$\Delta S_E(|v_{\vec{p}}|, |v_{\vec{k}})$ in Tree-level, for}$\lambda$=[0.5(top left), 1(top right), 2.5(bottom left), 3(bottom right)]}
\label{fig:contours2}
\end{figure}





\section{Conclusions and Perspectives}

We performed a detailed study of the variation of von-Neumann entanglement entropy in the elastic scattering
of a bipartite system  comprised of two interacting scalar fields, $A$ and $B$, to one loop order in perturbation theory. By constructing the final state and the reduced density matrix of the subsystem $A$ via S-matrix formalism in quantum field theory, we quantified the entanglement generated when we scatter a separable or an entangled momentum initial state. The correlations between the parties show explicit dependence on the speed (energy) between the colliding particles.

For an initially separable state, the velocity in the center of mass frame correspondent to the maximal entropy generation increases almost linearly with the coupling strength (consistently with perturbation theory) up to a certain value $\lambda=2$ (see fig. \ref{fig:velocitylinear}) for both tree and one loop correction. After that, there is a significant increment when one loop corrections are taken into account. It is noteworthy that the entropy increases by a factor of 10 as the coupling strength increases in the range $0.5$ to $3.0$ in our units, fig. \ref{fig:entropylinear}.

The maximum values achieved by the variation of the entanglement entropy is approximately 1.5 times higher when the initial state is entangled than when it is separable. It is also worth mentioning, that although we performed such calculations in the specific case  $\alpha_p = \alpha_k$, the maximum values achieved by the variation of the entanglement entropy remain unchanged for any other combination of non null coefficients. What does vary is the domain for which the same values are registered. When the coefficients for both states are the same, it can be witnessed in (fig.\ref{fig:contours2}) that the domain is somewhat symmetric. This is not the case if the coefficients are different, the domain of maximum variation extends to values along the velocity axis which happens to be associated with the state with the lower coefficient (this can be understood because such values are less pertinent to the outcome since they "weigh" less when calculating the variation, so they have a broader domain for which they produce the same value). This entails that there is a greater degree of possible manipulation for an initial entangled state such that it reproduces the same outcome of the values of the variation.

It would be interesting to study the entanglement generation in momenta/spin in final states of quantum electrodynamics processes, say a two-fermion scattering or a particle decay. Furthermore, the entanglement of a two particle state is generally not Lorentz invariant although certain states and partitions are. Such an analysis can be made in the light of Bell's inequality violation from first principles in a explicitly covariant framework of S-matrix quantum field  theoretical
calculation. Moreover, this may shed light on which entanglement measure should be considered to estimate quantum correlations as well as study the role of spin operators and superselection rules in entanglement generation. Such an analysis is also useful as a way of increasing cross sections, for instance in studying two photon scattering.


\bibliographystyle{plain}
\bibliography{references}

\begin{thebibliography}{}
\expandafter\ifx\csname url\endcsname\relax
  \def\url#1{\texttt{#1}}\fi
\expandafter\ifx\csname urlprefix\endcsname\relax\def\urlprefix{URL }\fi
\expandafter\ifx\csname href\endcsname\relax
  \def\href#1#2{#2} \def\path#1{#1}\fi

\end{thebibliography}


\section*{\bf References}
\vspace{0.5cm}

\begin{thebibliography}{9}

\bibitem{EPR} A. Einstein, B. Podolski and N. Rosen, Phys. Rev. {\bf{47}} (1935) 777.
\bibitem{Bell} J. Bell, ``On the Einstein Podoslky Rosen paradox" (1964) in: Speakable and unspeakable in quantum mechanics" Cambridge, Cambridge University Press, $2^{nd}$ edition, 2004.
\bibitem{Aspect} A. Aspect, P. Grangier and G. Roger, Phys. Rev. Lett. {\bf{49}} (1982) 91.
\bibitem{Hensen} B. Hensen et al., Nature {\bf{526}} (2015) 682.
\bibitem{CHSH} J. F. Clauser, M. A. Horne, A. Shimony and R. A. Holt, Phys. Rev. Lett. {\bf{23}} (1969) 880.
\bibitem{RQIA} C. H. Bennet et al., Phys. Rev. Lett. {\bf{83}} (1999) 3081; A. K. Ekert, Phys. Rev. Lett. {\bf{67}} (1991) 661; A. Peres, P. F. Scudo and D. R. Terno, Phys. Rev. Lett. {\bf{88}} (2002).
\bibitem{Friis} N. Friis, D. E. Bruschi, J. Louko and I. Fuentes, Phys. Rev. {\bf{D85}} (2012) 081701; N. Friis, M. Huber, I. Fuentes and D. E. Bruschi, Phys. Rev. {\bf{D86}} (2012) 105003.
\bibitem{Bertlmann} N. Friis, R. Bertlmann and M. Huber, Phys. Rev. {\bf{A81}} (2010) 042114.
\bibitem{Menicucci} E. Martin-Martinez and N. C. Menicucci, Class. Quant. Grav. {\bf{31}} (2014) 214001.
\bibitem{Nambu} Y. Nambu, Phys. Rev. {\bf{D78}} (2008) 044023.
\bibitem{Parentani} D. Campo and R. Parentani, Braz. J. Phys. {\bf{35}} (2005) 1074.
\bibitem{Gustavo} G. Souza, K. M. Fonseca-Romero, Marcos Sampaio and M. C. Nemes, Phys. Rev. {\bf{D90}} (2014) 125039.
\bibitem{BF} N. D. Birrell and L. H. Ford, Ann. of Phys. {\bf{122}} (1979) 1.
\bibitem{Fuentes0} J. L. Ball, I. Fuentes and F. P. Schuller, Phys. Lett. {\bf{A}} 359 (2006) 550.
\bibitem{Helder2} H. Alexander, Marcos Sampaio, Paul Mansfield, I. G. da Paz G. Souza, to appear in Eur. Lett. (2016).
\bibitem{Helder1} H. Alexander, Marcos Sampaio, Paul Mansfield and G. Souza, Eur. Lett. {\bf{111}}  (2015) 60001.
\bibitem{Fujikawa} K. Fujikawa, C. H. Oh and C. Zhang, Phys. Rev. {\bf{D90}} (2014) 025028.
\bibitem{Yongram} N. Yongram and E. Manoukian, Forsch. Phys. {\bf{61}} (2013) 668.
\bibitem{Seki1} S. Seki and S.-J. Sin, Phys. Lett.. {\bf{B735}} (2014) 272.
\bibitem{Pachos} J. Pachos and E. Solano, Quant. inf. and Comp. {\bf{3}} (2003) 115.
\bibitem{Seki2} S. Seki, I. Y. Park and S.-J. Sin, Phys. Lett. {\bf{B743}} (2015) 147.
\bibitem{Ratzel} D. Ratzel, M. Wilkens and R. Menzel, ``The effect of polarization entanglment in photon-photon scattering", hep-th/1605.00582v2.
\bibitem{LHC} D. d'Enterria and G. Silveira, Phys. Rev. Lett. {\bf{111}} (2013) 080405.

\end{thebibliography}

\end{document}